\documentclass[twocolumn,prl,aps,showpacs,superscriptaddress]{revtex4}

\usepackage{amssymb, amsmath, graphicx}
\usepackage[english]{babel}
\usepackage{bm}
\usepackage{graphicx}
\usepackage[latin1]{inputenc}

\long\def\symbolfootnote[#1]#2{\begingroup%
 \def\thefootnote{\fnsymbol{footnote}}\footnote[#1]{#2}\endgroup}
\long\def\symbolfootnotemark[#1]{\begingroup%
 \def\thefootnote{\fnsymbol{footnote}}\footnotemark[#1]\endgroup}

\begin{document}
\author{Marten Richter}
\email[]{marten.richter@tu-berlin.de}
\affiliation{Institut für Theoretische Physik, Nichtlineare Optik und
Quantenelektronik, Technische Universität Berlin, Hardenbergstr. 36, 10623
Berlin, Germany}
\author{Alexander Carmele}
\affiliation{Institut für Theoretische Physik, Nichtlineare Optik und
Quantenelektronik, Technische Universität Berlin, Hardenbergstr. 36, 10623
Berlin, Germany}
\author{Anna Sitek}
\affiliation{Institut für Theoretische Physik, Nichtlineare Optik und
Quantenelektronik, Technische Universität Berlin, Hardenbergstr. 36, 10623
Berlin, Germany}
\affiliation{Institute of Physics, Wroc{\l}aw University of Technology, Wybrze{\.z}e
Wyspia{\'n}skiego 27, 50-370 Wroc{\l}aw, Poland}
\author{Andreas Knorr}
\affiliation{Institut für Theoretische Physik, Nichtlineare Optik und
Quantenelektronik, Technische Universität Berlin, Hardenbergstr. 36, 10623
Berlin, Germany}

\title{Few-photons model of the optical emission of semiconductor quantum dots}
\begin{abstract}
The Jaynes-Cummings model provides a well established theoretical framework for single electron two level systems in a radiation field. Similar exactly solvable models for semiconductor light emitters such as quantum dots dominated by many particle interactions  are not known.
We access these systems by a generalized cluster expansion, the photon-probability-cluster-expansion: a  reliable approach for few photon dynamics in many body electron systems. As a first application, we discuss vacuum Rabi flopping and show that their amplitude determines the number of electrons in the quantum dot.
\end{abstract}
\maketitle
\date{\today}

\maketitle
{\it Introduction:} For describing the quantum dynamics of semiconductor light emitters  involving weak light-matter coupling and large photon numbers,  an expansion involving mean field quantities and their fluctuations (often called cluster or correlation expansion) provides a well controlled  theoretical scheme to treat many particle systems of electrons, phonons and photons \cite{Lindberg:PhysRevB:88,
Kira:ProgQuantumElectron:99,
Ahn:PhysRevB:05,
Gies:PhysRevA:07}.
However, for semiconductor emitters operated in the limit of few photons \cite{Michler:Nature:00}
 and few electronic levels, such as quantum dot-wetting  layer (QD-WL) systems (cf. Fig. \ref{scheme}a)) \cite{Gies:PhysRevA:07}, the cluster expansion (as illustrated below) breaks down, since it is based on the assumption that fluctuations of mean field quantities have minor influence.  With the rise of  quantum information and the search for proper solid state emitters, QD-WL systems for single and entangled photons are  in the focus of current research \cite{Yuan:Science:02,
Ahn:PhysRevB:05,Gies:PhysRevA:07}
.  Therefore, there is an urgent need for the extension of the standard cluster expansion.  Such an extension - the photon probability cluster expansion (PPCE) is  developed in this Letter.

 \begin{figure}[b]
\includegraphics[width=6cm]{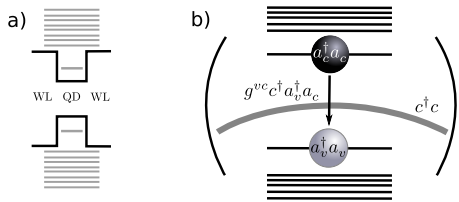}
\caption{ Scheme of the model system: a) a QD is embedded inside a WL leading to confined QD states coupled to a WL continuum, b) the emission of the confined states couples to a photon mode in a cavity.} 
\label{scheme}
\end{figure}
We generalize the cluster expansion \cite{Kira:ProgQuantumElectron:99,Ahn:PhysRevB:05,
Gies:PhysRevA:07} to the limit of few photons in systems with few quantum confined electronic levels such as semiconductor QD-WL systems, cf. Fig. \ref{scheme}a). The strength of our method is also to keep the accurate results of the standard cluster expansion for large photon numbers\cite{Kira:ProgQuantumElectron:99,Ahn:PhysRevB:05,
Gies:PhysRevA:07} and the strong coupling limit for high quality cavities \cite{Reithmaier:Nature:04}.\\\indent
To illustrate the standard cluster expansion and its break-down, we  introduce $c^\dagger$, $c$ as the Bosonic construction/destruction operators  for a single photon mode with frequency $\omega_0$ and $a^\dagger$, $a$ as the Fermionic operators for electrons.
Subscripts $c$, $v$, denote the lowest confined conduction and valence band state of the QD, respectively.
The corresponding model of the QD-WL system is depicted in Fig. \ref{scheme} b). In the following, this model will be used to illustrate our approach to reproduce exactly the Jaynes-Cummings model (JCM) (one electron approximation in atomic two level systems (i))  and extent it to many electron systems (QD-WL systems (ii)). To start, we illustrate the difference between the single and many electrons systems:
 The  Jaynes-Cummings model (JCM) \cite{Jaynes:ProcIEEE:63},
 containing {\it only one electron in an atomic two level system} is exactly solvable by expanding the solution into states having  distinct photon number and electronic state. Therefore, quantities of interest such as photon number $\langle c^\dagger c\rangle$, photon-photon correlation $\langle c^\dagger c^\dagger c c\rangle$, electronic densities $\langle a^\dagger_v a_v^{\phantom{\dagger}}\rangle$, $\langle a^\dagger_c a_c^{\phantom{\dagger}}\rangle$  can be calculated exactly. Here, the expectation value is denoted by $\langle \dots \rangle=\mathrm{tr}(\dots \rho)$, where $\rho$ is the statistical operator. In particular, by the exact expression $\langle a^\dagger_c a_v a^\dagger_v a_c \rangle=\langle a^\dagger_c  a_c \rangle$ for a single electron, the hierarchy of quantum correlations is closed\cite{Mandel::95}.   Unfortunately, however the JCM does not provide a scheme to include corrections resulting from further fermionic particles beyond the two level, one electron approximation.
Beyond the JCM, i.e. in many electron systems, such as an occupied WL, Fig. \ref{scheme},  non-factorizing correlations such as $\langle a^\dagger_c a^{\phantom{\dagger}}_v a^{\dagger}_v a^{\phantom{\dagger}}_c\rangle^c=\langle a^\dagger_c a_v a^\dagger_v a_c \rangle-\langle a^\dagger_c  a_c \rangle\neq0$ are important \cite{Gies:PhysRevA:07,Schneebeli:PhysRevLett:08}. 
{ This is the case in particular for  QDs embedded in a WL 
\cite{Gies:PhysRevA:07}, since in this case the electrons in the WL and in the QD  are Coulomb coupled: Electrons are scattered from the WL to QDs and vice versa, which lead to a correlation between electrons in the QDs and in 
WL\cite{Gies:PhysRevA:07}. The first correlation contribution requires an electron and a hole to be combined for emission.}
\\\indent
So far the standard cluster expansion of $\langle a^\dagger_c a^{\phantom{\dagger}}_v a^{\dagger}_v a^{\phantom{\dagger}}_c\rangle^c$ provides access only to large photon number dynamics or in weak coupling-short time limit ($\hbar |g^{vc}|\cdot t \ll 1$, where $g^{vc}$ is the light matter coupling constant, see below). This is, because the standard cluster expansion constitutes a hierarchy of equations for
correlations, in which the dynamics of relevant observables   are not connected directly to the  photon number state. Examples include quantities such as $\langle a^\dagger_c a^{\phantom{\dagger}}_c c^\dagger  c \rangle$, which include contribution from photon states with one and more photons
.
 This will be illustrated later using Eq. (\ref{cc}),
  it is clear that this kind of expansion is not  suitable, if correlations with a distinct number of photons are of crucial importance.

A theoretical scheme, which attacks this problem for a single photon mode, but is easy to extend to multiple photon modes, is presented in this Letter. This new approach will allow to treat few photon dynamics  for  quantum information in many particle systems such as QD-WL systems with few electronic levels, that contribute to the emission.

{\it Model system:}  We illustrate the proposed photon-probability cluster expansion (PPCE) using   the basic light-matter Hamiltonian $H=H_0+H_{el-pt}$\cite{Mandel::95}:
\begin{eqnarray}
H_0&=&\hbar \omega_0 c^\dagger c+\hbar \varepsilon_c a^\dagger_c a^{\phantom{\dagger}}_c+\hbar \varepsilon_v a^\dagger_v a^{\phantom{\dagger}}_v\label{ham1}\\
H_{el-pt}&=&-\hbar(g^{vc}a^\dagger_v a_c c^\dagger+{g^{vc}}^* a^\dagger_c a_v c),\label{ham3}
\end{eqnarray}
 (illustrated in Fig.\ref{scheme}b)).
 For simplicity, we consider only a valence band $v$ and conduction band $c$ state with the energies $\hbar \varepsilon_v$ and $\hbar \varepsilon_c$ respectively. The off-diagonal coupling matrix  $g^{vc}$ denotes the electron-photon-interaction strength. Additionally,  the two level QD is assumed to be embedded,  in an electronically occupied WL. In such a many particle system   the JCM-valid factorization $\langle a^\dagger_v a_c^{\phantom{\dagger}} a_c^\dagger a_v^{\phantom{\dagger}}\rangle=\langle a^\dagger_v a_v^{\phantom{\dagger}}\rangle$ has to be replaced by \cite{Kira:ProgQuantumElectron:99,
Gies:PhysRevA:07}  $\langle a^\dagger_v a_c^{\phantom{\dagger}} a_c^\dagger a_v^{\phantom{\dagger}}\rangle=\langle a^\dagger_v a_v^{\phantom{\dagger}}\rangle(1-\langle a^\dagger_c a_c^{\phantom{\dagger}}\rangle)+\langle a^\dagger_v a_c^{\phantom{\dagger}} a_c^\dagger a_v^{\phantom{\dagger}}\rangle^{\tilde{c}}$
\footnote{The $\tilde{}$ on the index $c$ is used to distingish this correlation from the earlier introduced
 $\langle a^\dagger_c a_v  a^{\dagger}_v a_c\rangle^c$.}
. Clearly, because  the one electron assumption in the QD-WL system fails,  Pauli-blocking terms are introduced  \cite{Gies:PhysRevA:07}. Since the correlation $\langle a^\dagger_v a_c^{\phantom{\dagger}} a_c^\dagger a_v^{\phantom{\dagger}}\rangle^{\tilde{c}}$ cannot be treated exactly, an analytical solution of the model is not possible anymore, and the cluster expansion of $\langle a^\dagger_v a_c^{\phantom{\dagger}} a_c^\dagger a_v^{\phantom{\dagger}}\rangle$ provides only  a perturbational approach. 
\begin{figure}
\includegraphics[clip,width=6cm]{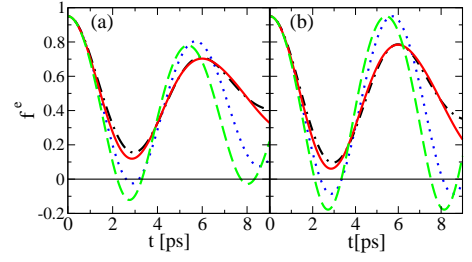}
\caption{ The JCM (solid/red), standard cluster-expansion atom like using one electron assumption (dashed/green), and standard cluster-expansion using $f^e_{sp}=f^e f^h$ in QD-WL system (dotted/blue) and PPCE using $f^e_{sp,n}=f^e_n f^h_n$ in QD-WL system in new scheme (dotted dashed/black). With the initial conditions: $f^e=f^h=0.95$, $p_0=0.8$, $p_1=0.2$ and a coupling constant $|g^{vc}|=0.5~1/\mathrm{fs}$ for a QD resonant with the cavity mode. ( (a)$\gamma_p=1/(10\mathrm{ps})$ and (b)$\gamma_p=0/(\mathrm{ps})$)} 
\label{alexcomp}
\end{figure}

We start showing that   the cluster expansion  yields   unphysical behaviour in the few photon  strong coupling limit for two electronic levels: Up to the order $|g^{vc}|^2$, using the Heisenberg equation of motion, the cluster expansion reads \cite{Ahn:PhysRevB:05,Gies:PhysRevA:07}:
\begin{eqnarray}
 \partial_t\langle c^\dagger c\rangle&=&-\partial_t f^e=-\partial_t f^h=-2\mathrm{Im}(g^{vc*}\langle a^\dagger_v a_c^{\phantom{\dagger}}c^\dagger \rangle)\\
\partial_t \langle a^\dagger_v a_c^{\phantom{\dagger}}c^\dagger \rangle &=&\imath (\varepsilon_v-\varepsilon_c+\omega_0+\imath \gamma_p) \langle a^\dagger_v a_c^{\phantom{\dagger}}c^\dagger \rangle-\imath g^{vc*} f^e_{sp}\nonumber\\
&&-\imath g^{vc*}\langle c^\dagger c\rangle (f^e+f^h-1).
\end{eqnarray}
Here, we introduced the hole occupation  $f^h=1-\langle a^\dagger_v a^{\phantom{\dagger}}_v\rangle$, the electron occupation $f^e=\langle  a^\dagger_c a^{\phantom{\dagger}}_c\rangle$ and the phenomenological pure dephasing constant $\gamma_p$. The spontaneous emission term $f^e_{sp}$ is equal to $f^e$ in case of the JCM  and $f^e_{sp}=\langle a^\dagger_v a_c^{\phantom{\dagger}} a_c^\dagger a_v^{\phantom{\dagger}}\rangle=f^e f^h$ for the QD-WL case \cite{Gies:PhysRevA:07}. 
In  Fig. \ref{alexcomp} we plot the dynamics of the electron occupation for: (i)  exact JCM model \cite{Jaynes:ProcIEEE:63}, (ii) cluster-expansion using one electron assumption $f^e_{sp}=f^e$, (iii) for the QD-WL system $f^e_{sp}=f^e f^h$ and also (iv) for the below developed PPCE,  introduced later.

 The exact JCM solution for the one electron assumption  leads to  well known  vacuum Rabi flops.
Fig. \ref{alexcomp} shows clearly  that the standard cluster expansion of the JCM and the QD-WL system   lead to  negative and unphysical results in the few photon  limit.  However, we note that in the many QD  and/or many photon case $\langle c^\dagger c\rangle\gg 1$,  correlation expansion provides an excellent tool \cite{Gies:PhysRevA:07,Schneebeli:PhysRevLett:08}.
As expected, all approaches agree for small times (weak coupling limit $\hbar^{-1} |g^{vc}| t\ll 1$).  

 The breakdown of the cluster expansion results from the fact that the system is completely characterized with mean single particle quantities. This is problematic in the limit of few electrons and photons, since correlations (in particular fluctuations) between electrons and photon number states are crucial in this limit.  To improve  this situation, we focus now on the development of the PPCE.

{\it Photon-Probability Cluster Expansion (PPCE):} Important quantities to characterize a quantum optical field are the normalized intensity-intensity correlation function $g^{(2)}(\tau=0)=\frac{\langle c^\dagger c^\dagger c c \rangle}{\langle c^\dagger c \rangle^2}$ and the photon intensity $g^{(1)}(\tau=0)=\langle c^\dagger c \rangle$\cite{Mandel::95}. Both quantities can be measured and they are directly connected to the photon number $\langle c^\dagger c\rangle$ and  intensity-intensity expectation value $\langle c^\dagger c^\dagger c c\rangle$. 
 The idea - to circumvent the problems characteristic for the cluster expansion (cf. Fig. \ref{alexcomp}) - is to   formulate all quantities in terms of  $n$ photon probability $p_n=\big\langle |n\rangle \langle n |\big\rangle$, where  $|n\rangle$ denotes the Fock-state of $n$ photons in the system. This idea is already succesfully applied in JCM \cite{Jaynes:ProcIEEE:63}, but needs to be generalized to the many electron case. The PPCE introduced now combines the advantages of the exact JCM-solution and the standard cluster expansion.  The observables read in the Fock basis:
\begin{eqnarray}
\langle c^\dagger  c\rangle&=&
\sum_{n=1}^\infty n p_n,\;\label{cc}
\langle c^\dagger c^\dagger c c\rangle
=\sum_{n=2}^\infty n(n-1) p_n.\label{cccc}
\end{eqnarray}
Eq. (\ref{cc}) show that
 it is sufficient to calculate the equations of motion for the probability to find $n$ photons $p_n=\big\langle |n\rangle \langle n |\big\rangle$ in the system.
 Heisenberg equations  for $p_n$, using the Hamiltonian,  Eqs. (\ref{ham1}-\ref{ham3}), couple to assisted occupation probabilities: $\big\langle |n\rangle \langle n |a^\dagger_c a^{\phantom{\dagger}}_c\big\rangle$, $\big\langle |n\rangle \langle n |a^\dagger_v a^{\phantom{\dagger}}_v\big\rangle$ and transitions $\big\langle |n\rangle \langle n+1 |a^\dagger_c a^{\phantom{\dagger}}_v\big\rangle$ or similar quantities if a more general system is considered. We discuss the application of the theory to the JCM (one electron) and the QD-WL-system (many electrons) separately:

For the case of   one electron ({\it JCM case}),
 the equation of motion for the probability $p_n$   reads:
\begin{eqnarray}
 \partial_t p_n&=&-2\sqrt{n} \mathrm{Im}\big(g^{vc} \big\langle |n\rangle \langle n-1| a^\dagger_v a^{\phantom{\dagger}}_c\big\rangle\big)\nonumber\\&&+2\sqrt{n+1} \mathrm{Im}\big({g^{vc}} \big\langle |n+1\rangle \langle n| a^\dagger_v a^{\phantom{\dagger}}_c\big\rangle\big).\label{jcm_pn}
\end{eqnarray}
The $n$-photon probability $p_n$ is driven by the photon transition assisted polarization $\big\langle |n+1\rangle \langle n| a^\dagger_v a^{\phantom{\dagger}}_c\big\rangle$:
\begin{eqnarray}
 &\partial_t& \big\langle |n+1\rangle \langle n| a^\dagger_v a^{\phantom{\dagger}}_c\big\rangle\nonumber\\&&=\imath (\varepsilon_v-\varepsilon_c+\omega_0+\imath \gamma_p)\big\langle |n+1\rangle \langle n| a^\dagger_v a^{\phantom{\dagger}}_c\big\rangle\nonumber\\
&&+\imath {g^{vc}}^*\sqrt{n+1}(p_{n+1}-f^h_{n+1})
-\imath {g^{vc}}^*\sqrt{n+1}f^e_{n}.\label{trans_jc}
\end{eqnarray}
Besides the free energy rotation quantum optical absorption and emission processes occur in the last line.
 These processes are related to $f^h_{n}=p_n-\big\langle |n\rangle \langle n | a^\dagger_v a^{\phantom{\dagger}}_v\big\rangle$ and $f^e_n=\big\langle |n\rangle \langle n | a^\dagger_c a^{\phantom{\dagger}}_c\big\rangle$ for hole and electron densities  assisted by   $n$ photons.
Finally, the electron and hole densities $f^{e/h}_n$ couple to quantum optical transitions:
\begin{eqnarray}
\partial_t f^{e/h}_n&=&2\sqrt{n+1} \mathrm{Im}\big({g^{vc}} \big\langle |n+1\rangle \langle n| a^\dagger_v a^{\phantom{\dagger}}_c\big\rangle\big).\label{den_eq}
\end{eqnarray}
To close the hierarchy of equations of motion we used $\big\langle a^\dagger_i a^\dagger_j a_k a_l|n\rangle \langle n|\big\rangle=0$ strictly valid only for the single electron case, since two electrons are annihilated.
Eqs.(\ref{jcm_pn}-\ref{den_eq}) without pure dephasing reproduces the JCM \cite{Jaynes:ProcIEEE:63}. This is an important benchmark of the new PPCE.

If we consider more than one electron ({\it QD-WL case}), $\partial_t \big\langle |n+1\rangle \langle n| a^\dagger_v a^{\phantom{\dagger}}_c\big\rangle$, couples to $\big\langle a^\dagger_v a_c^\dagger  a_c^{\phantom{\dagger}} a_v^{\phantom{\dagger}}|n\rangle \langle n|\big\rangle\neq0$.
 In order to truncate the PPCE hierarchy of equations of motion at least approximately, we need factorization rules for correlation between electrons, but also for other bosons  to e.g. introduce cavity loss and radiative decay consistently.

We use and generalize two basic factorization limits, Hartree-Fock and Born approximation (i,ii), respectively.\\
{\it (i) The Hartree-Fock approximation}  $\langle a^\dagger_i a^\dagger_j a_k a_l \rangle\approx\langle a^\dagger_i a_l \rangle \langle a^\dagger_j a_k\rangle- \langle a^\dagger_i a_k \rangle \langle a^\dagger_j a_l\rangle$ is based on the assumption \cite{Lindberg:PhysRevB:88,Fick::90}, that the density matrix $\rho$ can  be  described  by single particle quantities using the Generalized Canonical  Statistical Operator (GCSO) $\rho(t) \approx \mathrm{exp}[-{\sum_{ij} \lambda_{ij}(t)a^\dagger_i a_j^{\phantom{\dagger}}}/({k_B T})]/Z$ \cite{Fick::90} with the Lagrange parameters $\lambda_{ij}(t)$.
For the n-photon number contribution of the density matrix  $\langle n|\rho(t)|n\rangle/p_n$,
 a photon number dependence of the Lagrange parameters is introduced:   $\langle n|\rho(t)|n\rangle/p_n
 \approx \mathrm{exp}[-{\sum_{ij} \lambda_{ij}^n(t)a^\dagger_i a_j^{\phantom{\dagger}}}/({k_B T})]/Z$.  From this, we recognize the factorization rule:
 $\big\langle|n\rangle\langle n| a^\dagger_1 a^\dagger_2 a_3 a_4 \big\rangle= \big(\big\langle|n\rangle\langle n| a^\dagger_1 a_4 \big\rangle \big\langle|n\rangle\langle n| a^\dagger_2 a_3\big\rangle- \big\langle|n\rangle\langle n|a^\dagger_1 a_3 \big\rangle \big\langle|n\rangle\langle n| a^\dagger_2 a_4\big\rangle\big)/p_n
$. \\
{\it (ii) The second order-Born approximation}
for a correlated electron-photon system embedded in a bosonic bath (e.g. phonons or other photon modes $b^\dagger$, $b$)  assumes that  the density matrix can be  factorized: $\rho\approx \rho_B\otimes \rho_{el-pt}$, which implies: $\big\langle|n\rangle\langle n| a^\dagger_1  a_2  b^\dagger_3 b_4\big\rangle\approx\big\langle|n\rangle\langle n| a^\dagger_1  a_2\big\rangle \langle  b^\dagger_3 b_4\rangle$.
In particular, for QD-WL system, to some electronic states  bath like properties (e.g. WL states) can be attributed, so  a factorizing density matrix $\rho=\rho_{el-uncorr}\otimes\rho_{ph+corr}$ can be assumed. This leads to a factorization rule: $\big\langle |n\rangle \langle n | a^\dagger_i a_j\big\rangle\approx \big\langle |n\rangle \langle n |\big\rangle \langle a^\dagger_i a_j \rangle$ for WL states $i,j$.\\
\indent Our factorization rules can be extended to include high order PPCE correlations, preserving the important electron-photon number correlations.
We now evaluate the equations of motion in the PPCE approach:
In comparison to the single electron case,
the equation for $p_n$ (Eq. (\ref{jcm_pn})) does not change.
Eq. (\ref{trans_jc}) is modified for the QD-WL system case  using the modified Hartree-Fock factorization rule:
\begin{eqnarray}
 &\partial_t& \big\langle |n+1\rangle \langle n| a^\dagger_v a^{\phantom{\dagger}}_c\big\rangle\nonumber\\&&=\imath (\varepsilon_v-\varepsilon_c+\omega_0+\imath \gamma_p)\big\langle |n+1\rangle \langle n| a^\dagger_v a^{\phantom{\dagger}}_c\big\rangle\nonumber\\&&
+
\imath g^{vc}\sqrt{n+1}{(p_{n+1}-f^h_{n+1})(p_{n+1}-f^e_{n+1})}/{p_{n+1}}\nonumber\\&&
-\imath g^{vc}\sqrt{n+1}{f^h_{n} f^e_{n}}/{p_n}.\label{PPCE1}
\end{eqnarray}
In comparison  to the one electron assumption, Eq. (\ref{trans_jc})  the terms $(p_{n+1}-f^h_{n+1})$ and $f^e_{n}$   are replaced with $(p_{n+1}-f^h_{n+1})(p_{n+1}-f^e_{n+1})/p_{n+1}$ and $f^h_{n} f^e_{n}/p_n$ respectively. 
Similarly, Eq. (\ref{den_eq}) does not change. { (In Eq. (\ref{PPCE1}), some sourceless terms  including $\big\langle |n\rangle \langle n | a^\dagger_v a_c\big\rangle$  are neglected.)}  
Eqs. (\ref{jcm_pn}),(\ref{den_eq}),(\ref{PPCE1}) constitute the new quantum optical dynamics for a few photon/ few electronic level system.

\begin{figure}
\includegraphics[width=4cm]{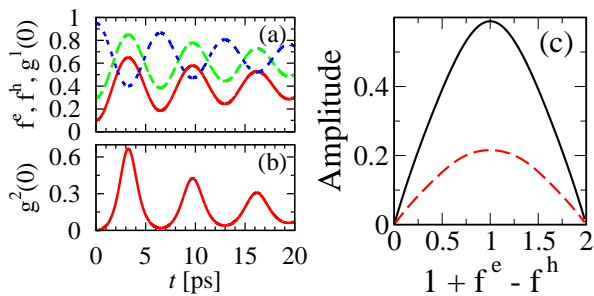}
\includegraphics[clip,width=3.7cm]{fig3c.eps}
\caption{ The PPCE solution for semiconductor QDs. With the initial conditions: $f^h=0.3$ and $f^e=0.1$, $p_0=0.05$, $p_1=0.95$ and a coupling constant $g=0.5~ \mathrm{meV}$ for a QD resonant with the cavity mode and  $\gamma_p=(10 \mathrm{ps})^{-1}$. In (a) the different lines are $f^e$ (solid/red), $f^h$ (green/long dashed), $g^1(0)$ (blue/short dashed). (c) The maximum value of $g^{(2)}$ for PPCE is plotted over the number of electrons in the QD, with $p_1=1.0$ (black/straight) and $p_2=1.0$ (red/dashed).} 
\label{semicond}
\label{g2max}
\end{figure}
{\it Discussion:}
The best benchmark for the new PPCE scheme is the case of vacuum Rabi flops where quantum fluctuations dominate the dynamics.
{ For strong coupling, the in-scattering rate from the WL has to be insignificant compared to $g^{vc}$. Ref. \onlinecite{Nielsen:PhysRevB:04a} suggest for $g^{vc}$ chosen here, WL densities $\ll 10^{11} \mathrm{cm}^{-2}$.}
{ Now,} 
we reiterate our discussion of Fig. \ref{alexcomp}. Obviously, only  the PPCE (dotted dashed) provides physical reasonable results as opposed to the standard cluster expansion method, since it guarantees positivity of the electronic density. In particular, we recover  the exact JCM solution. The Pauli blocking term occuring in the QD-WL system  does not lead to strong deviations for not long  times from the  JCM case.
 For the calculations in Fig. \ref {alexcomp}, we assumed, that initially there is  one electron $(f^e+(1-f^h))=1$ in the QD $f^h=f^e$. 
For the numerically evaluation, we choose as initial correlations  a finite photon number distribution $p_n(t_0)$,  whereas the initial transitions $\big\langle |n+1\rangle \langle n| a^\dagger_v a^{\phantom{\dagger}}_c\big\rangle$ will be set to zero, i.e. assuming an uncorrelated initial state. 
  Also vanishing initial correlations between electron and hole system are assumed: The photon number, electron and hole densities will be set to $f^h_n=f^h\cdot p_n(t_0)$ and $f^e_n=f^e\cdot p_n(t_0)$. 
Comparing the PPCE  calculation  to the cluster expansion and the  JCM, only the 
 PPCE recovers the exact JCM result. Therefore, we assume that it is more valid for the QD-WL case than the standard cluster expansion.
{ The damping $\gamma_p$ of Rabi flopping  describes dephasing due to electron-phonon coupling and Coulomb correlations betweeen QD and WL \cite{Nielsen:PhysRevB:04a}.}

We now focus on the specific QD-WL properties. In Fig. \ref{semicond}, the dynamics of $f^e$, $f^h$, $g^{(1)}$ and $g^{(2)}$ is plotted for  semiconductor specific initial values $f^h\neq f^e$, occuring in QD-WL systems \cite{Gies:PhysRevA:07}. 
The average photon number oscillates between $0$ and $1$, the  $g^{(2)}$ between $0$ and $0.7$ is in the anti bunching regime \cite{Mandel::95}. After the first Rabi flop the system goes into the single photon emitter regime with a $g^{(2)}<0.5$\cite{Lounis:RepProgPhys:05}. 
{\it In contrast to the JCM, the strength of the vacuum Rabi flopping is reduced in dependence on how strong the electron and hole density deviates from the one electron assumption $f^h=f^e$. 
This implies that the amplitude of the Rabi flops might be used as a measure for the number of electrons in the actual QD.}
Since the amplitude of the Rabi flopping is directly connected to  $g^{(2)}$, it may be possible to stear $g^{(2)}$ with the deviation from the one electron assumption.
This is  illustrated more clearly in Fig. \ref{g2max}c), where the number of electrons in the  QD is plotted as a function of the maximum of $g^{(2)}$, which is directly connected to the maximum of vacuum Rabi flopping. It is seen that a maximum of $g^{(2)}$ is achieved, if we have only one electron $f_e= f_h$. This behaviour can be attributed to the Pauli-blocking terms $f_e^n\cdot f_h^n$, because  they drive the emission and are enhanced in the one electron case.  

{\it In conclusion:} a modified cluster expansion scheme to describe, semiconductor quantum dot - wetting layer quantum optical  devices in the strong coupling limit involving few photons  and few dominant electronic states  is introduced. This photon-probability-cluster expansion (PPCE), reproduces the Jaynes-Cummings model (JCM) as a well known benchmark in quantum electrodynamics, the standard cluster expansion cannot.
For the QD-WL system, we have shown that the amplitude of vacuum Rabi flopping and the maximum of $g^{(2)}$ depend on the numbers of electron and holes pumped into the quantum dot device. This is a first important prediction from the PPCE.

We envision, that the new scheme can be applied  to many solid state structures  with few electronic states limit in the few photon limit. The approach will stimulate further investigation beyond JCM: Our approach allows  to include electron-phonon interaction and Coulomb interaction additionally to the already demonstrated electron-photon interaction in a similar fashion.

We acknowledge support from Deutsche Forschungsgemeinschaft through SFB 787 and A.S. acknowledges financial support from the DAAD. 


\end{document}